\documentclass[universe,review,accept,moreauthors,10pt,a4paper]{mdpi} 
\firstpage{1} 
\articlenumber{x}
\doinum{10.3390/------}
\pubvolume{xx}
\pubyear{2017}
\copyrightyear{2017}
\externaleditor{Academic Editor: name}
\history{Received: date; Accepted: date; Published: date}

\Title{Microlensing and Its Degeneracy Breakers: Parallax, Finite Source, High-resolution Imaging, and Astrometry}

\Author{Chien-Hsiu Lee $^{1,\dagger}$}

\AuthorNames{Chien-Hsiu Lee}

\address[1]{%
$^{1}$ \quad Subaru Telescope, National Astronomical Observatory of Japan, 650 N Aohoku Pl. Hilo, HI 96720, USA; leech@naoj.org}

\corres{Correspondence: leech@naoj.org; Tel.: +1-808-934-7788}

\abstract{First proposed by Paczynski in 1986, microlensing has been instrumental in the search of compact dark matter as well as discovery and characterization of exoplanets. In this article, we provide a brief history of microlensing, especially on the discoveries of compact objects and exoplanets. We then review the basics of microlensing and how astrometry can help break the degeneracy, providing a more robust determination of the nature of the microlensing events. We also outline prospects that will be made by on-going and forth-coming experiments/observatories.}


\begin{document}



\section{Introduction}
According to Einstein’s theory of general relativity [1], massive foreground objects can induce strong space-time curvature, serving as gravitational lenses and focus the lights of background sources into multiple and magnified images that projected along the observer’s line-of-sight. Such gravitational lensing systems provide us unique opportunities to study dark matter that hardly reveal their existences via electromagnetic radiations, or very faint objects that are beyond the sensitivity of state-of-the-art instruments. However, based on the calculations of Chwolson [2] and Einstein himself ([3], upon the request of R. W. Mandl), if the foreground object is as compact and light as a stellar object, the chance of gravitational lensing is very slim and given the telescopes and instruments in the early 20th century, it is unlikely to observe such an event. 

	The situation has been changed with the advent of modern CCDs and wide-field surveys. Paczyński [4] first envisioned the search of Galactic dark matter in compact form using gravitational lensing method; as the angular separation of the lensed images are in the order of micro arcseconds, such phenomena are often called microlensing. In his calculations, Paczyński showed that the chance (or optical depth) of an massive object in the Galactic halo to serve as a lens and magnify a background star in nearby galaxy is 10-6. While Paczyńsk’s calculation confirmed Chwolson and Einstein’s speculations, it also suggested that based on modern instruments, we will be able to catch such microlensing events if monitoring a dense stellar field with more than a million stars at once. Trigged by Paczyński’s seminal paper, several experiments such as MACHO[5], Expérience pour la Recherche d'Objets Sombres (EROS[6]), Optical Gravitational Lensing Experiment (OGLE[7]), Microlensing Observations in Astrophysics (MOA,[8]) were carried out, aiming at the closest dense stellar fields — Magellanic Clouds. The first microlensing events were announced by the MACHO team [9], EROS team [10], and OGLE team [11]. After the first detections, the MACHO team continuously observed the Magellanic Clouds until the devastating bush fire destroyed the 50-inch Great Melbourne Telescope in January 2003. With the 5.7 years of survey data, they identify 13 microlensing events towards LMC [12], albeit 3 of them are likely contaminations from variable stars [13]. Nevertheless, with 10 microlensing events in hand, Bennett [13] concluded that 16\% of the Galactic halo is composed of massive compact objects with masses between 0.1 and 1 solar mass. On the other hand, with data gathered between  1990-1995 (EROS-1) and 1996-2003 (EROS-2), the EROS team concluded that less than 8\% of the Milky Way halo is consisted of massive compact objects with an average mass of 0.4 solar mass; they also further ruled out massive compact objects with masses between 0.6 x 10-7 and 15 solar masses to be the major component of the Milky Way halo [14]. Using the data gathered by OGLE in 1998-2000 (OGLE-II) and 2002-2009 (OGLE-III), Wyrzykowski et al. [15-17] also concluded that microlensing events towards both Magellanic Clouds can be well explained by self-lensing (both the lens and source are stars in the Magellanic Clouds) without invoking compact dark matter. Even more so, Besla et al. [18] presented studies of tidal streams between LMC and SMC, and showed that microlensing signals can be reproduced by stars in the stream. On the other hand, re-analysis of OGLE and EROS data [19] showed that some of the OGLE events can be caused by compact dark matter.

	The inconclusive results might originate from some draw backs of using LMC/SMC as source field. For example, towards LMC/SMC we are limited to a single line-of-sight of the Galactic halo due to our fixed position in the Milky Way. In addition, because of the proximity of Magellanic Clouds, their 3-D structures are non-negligible and the self-lensing rate is unknown. In this regard, Crotts [20], Baillon et al. [21], and Jeter [22] have proposed to use M31 as an alternative stellar field for microlensing searchers. First of all, at a distance of 770 kpc, the geometric effect is negligible compared to LMC/SMC. Secondly, we can probe different sight-lines towards M31, using either its bulge or disk as sources. In addition, besides the Milky Way halo, we can also probe the halo of M31. Tomaney \& Crotts [23] were the first ones to conduct M31 microlensing searches; they utilized the 1.8m Vatican Advanced Technology Telescope (VATT) on Mount Graham, and the 4m Mayall telescope on Kitt Peak, both in Arizona, USA, to observe M31 between 1994 and 1995 and presented six microlensing events in M31 [24]. They continuously monitored M31 with VATT and the 1.3m Michigan-Dartmouth-MIT telescope on Kitt Peak from 1997 till 1999 and reported 4 microlensing events [25]. At the mean time, Ansari et al. [26] carried out the Andromeda Gravitational Amplification Pixel Experiment (AGAPE) using the 2m Bernard Lyot Telescope at Pic du Midi de Bigorre observatory in the French Pyrenees in 1994 and 1995, finding one bright and short microlensing event. The successor of AGAPE, the Pixel-lensing Observations with the Isaac Newton Telescope-Andromeda Galaxy Amplified Pixels Experiment (POINT-AGAPE) made use of the Wide Field Camera mounted on the 2.5-m Isaac Newton Telescope and monitored two 33 x 33 arc minutes fields north and south of the M31 bulge. With data gathered from 1999 till 2001, Auriere et al. [27] first announced 1 microlensing event, followed by three more by Paulin-Henriksson et al. [28] and another three more by Calchi Novati et al. [29]. The full POINT-AGAPE were analyzed by three working groups based at Cambridge, Zurich, and London, leading to three [30], six [31], and ten events [32] , respectively. Using the very same data-set, the Microlensing Exploration of the Galaxy and Andromeda (MEGA) presented 14 microlensing events [33]. At the mean time, the Nainital Microlensing Survey (NMS) monitored M31 from 1998 till 2002 with the 1.04m Sampurnanand Telescope in India and presented 1 microlensing events [34]. The Pixel Lensing Andromeda collaboration (PLAN) also carried out M31 observations in 2007 using the 1.5m Loiano telescope in Italy and presented 2 microlensing events [35]. PLAN further extended M31 observations in 2010 using the 2m Himalayan Chandra Telescope (HCT), which resulted in another microlensing event [36]. In the mean time, the Wendelstein Calar Alto Pixellensing Project (WeCAPP) presented 12 microlensing from M31 data gathered by the 0.8-m telescope at Wendelstein observatory in Bavarian Alpes and the 2.2-m telescope at Calar Alto observatory in Spain between 1997 and 2008 [37]. While some of the reported events could be attributed to variables, and with the small number of reported events, it is hard to put a stringent constraint on the fraction of compact dark matter in the halo of M31. Nevertheless, with two bright events, i.e. POINT-AGAPE-S3/WeCAPP-GL1 [28,38]  and OAB-N2 [39] are hard to reconcile with self-lensing scenario and point toward the existence of compact dark matter in the halo of M31. The main draw back of these M31 microlensing searches is the very limited field-of-view of the detectors. With the advent of wide-field imager mounted on Pan-STARRS 1, which already yielded 6 microlensing events [40], we will be able to gather a large sample of M31 with different lines-of-sights towards the bulge and the disk component at the same time. This will hopefully pin down the compact dark matter fraction in the M31 halo in the near future.

While the dark matter search remains inconclusive, microlensing also reveals intriguing binary/planetary objects. On the theoretical side, Mao \& Paczynski [41] first suggested that microlensing can be used to search for binary and planetary companion; they also suggested that about 10\% of the microlensing events could originate from binaries. Gould \& Loeb [42] provided further theoretical investigations explicitly on identifying exoplanets using the microlensing methods. They found out that planets in solar-like systems will induce significant deviations from the single lens light curve in 10\% of the microlensing events found in the Milky Way. They also note that the planetary light curve signature will only last for about a day, which is relatively short compared to the ~ 1 month single lens event time scale, thus dedicated and high-cadence follow-up is essential to the discovery of planetary microlensing events. On the observation side, the first discoveries of binary microlensing was reported by Udalski et al. [43]. There were some candidates planetary microlensing events, e.g. MACHO 1997-BLG-041 [44], but further analysis with more follow-up data indicated that these can be better explained by binary events [45-46]. The first definitive planetary microlensing event is OGLE-2003-BLG-235/MOA-2003-BLG-53, as reported by Bond et al. [47]. Thanks to the dense light curve sampling by combining both OGLE and MOA photometry, Bond et al. [47] inferred a 1.5 Jupiter-mass planet in an orbit of 3 AU, if it is associated with a star on the main-sequence. Numerous planet systems have been identified since then, e.g. super Earth events like OGLE-2005-BLG-390Lb (5.5 Earth-mass [48]) or MOA-2007-BLG-192Lb (3.3 Earth-mass [49]), or analogs of Jupiter/Saturn like OGLE-2006-BLG-109 [50], to name a few. In addition, microlensing is also very powerful to detect free-floating planets. For example, Sumi et al. [51] identified 10 unbounded Jupiter-mass events with Einstein time scale shorter than 2 days from the 2006-2007 MOA data, suggesting these objects are twice as common as main-sequence stars in the Milky Way.
        Though there are numerous microlensing events detected up-to-date, most of the information we can extract from the microlensing light curve is the event time scale, which is a combination of the lens mass, proper motion, and distance. In the next section, we will review the basics of microlensing, both on the photometric and astrometric aspects. We will then show in Section 3 how to break the degeneracy with astrometry, followed by prospects in Section 4.

\section{Microlensing formalism}
When a mass (e.g. a brown dwarf, a star, a black hole, or a galaxy) passes between the observer and a background source (e.g. a star, a quasar, or a galaxy), the mass induces space-time curvature and serves as a ‘gravitational’ lens. The light rays from the background source are thus deflected; instead of observing the original source, the observer sees two distorted images (if the source is extended) projected onto the source plane (assuming a single, point-like mass). The position of the images on the source plane can be derived from the lens equation, as shown in Fig. 1. 

\begin{figure}[H]
\centering
\includegraphics[width=15 cm]{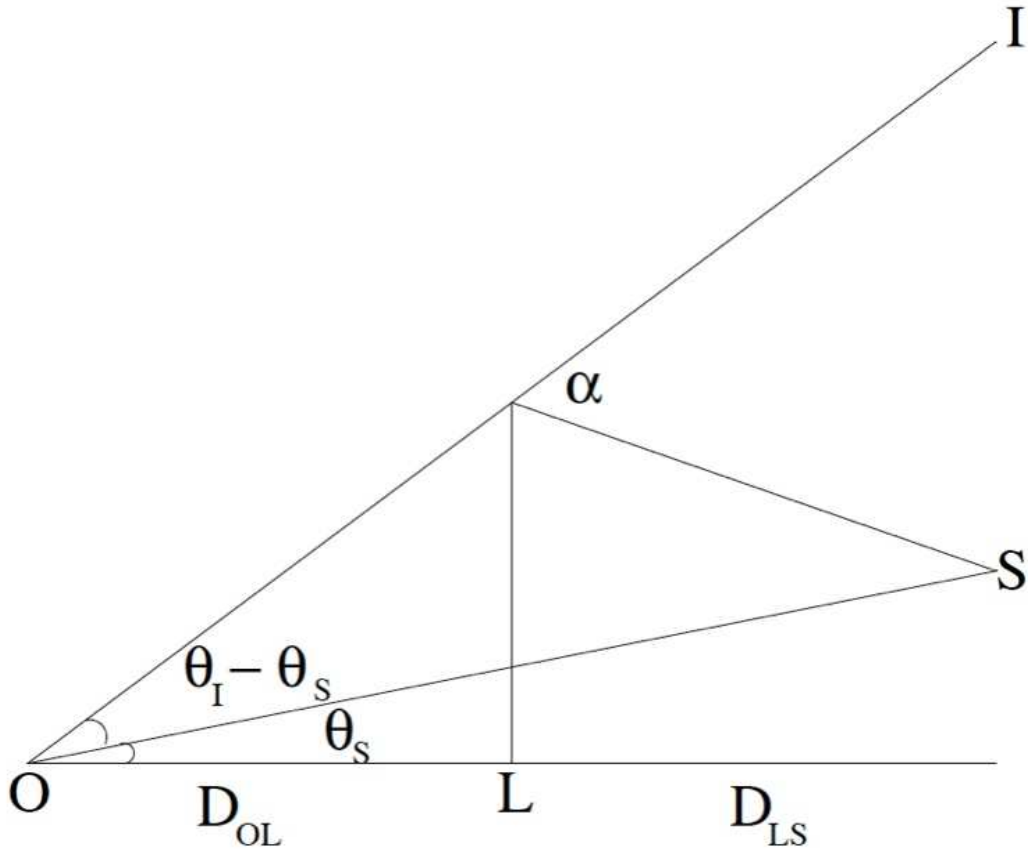}
\caption{Light paths of gravitational lensing.}
\end{figure}

In the triangle OIS, IS = $\alpha D_{LS} = (\theta_I-\theta_S)D_{OS}$, where $D_{OS} = D_{OL} + D_{LS}$ is the distance between the observer and the background source. The light bending angle
\begin{equation}
\alpha=\frac{4GM}{c^2D_{OL}\theta_I}
  \end{equation}

can be calculated from Einstein’s general theory of relativity [52]. Thus one can derive the lens equation
\begin{equation}
\theta_I(\theta_I-\theta_S)=\frac{4GM}{c^2}\frac{D_{LS}}{D_{OL}D_{OS}}:=\theta_E
\end{equation}

When the observer, the lens and the source are so well aligned that the lens overlaps with the source in the line-of-sight direction, the distorted images become a ring, a.k.a. the ‘Einstein Ring’. The angular Einstein Ring radius can be expressed as

\begin{equation}
\theta_E=\sqrt{\frac{4GM}{c^2}\big(\frac{1}{D_{OL}}-\frac{1}{D_{OS}}\big)}=0.902mas\big(\frac{M}{M_\odot}\big)^{1/2}\big(\frac{10kpc}{D_{OL}}\big)^{1/2}\big(1-\frac{D_{OL}}{D_{OS}}\big)^{1/2}
\end{equation}

in case of a point mass and a point source. Microlensing bears the name of ’micro’ because such events were first observed towards distant quasars, and the size of the Einstein Ring is in the scale of micro-arcsecond. If we normalize the length scale to $\theta_E$ and define
\begin{equation}
  u=\frac{\theta_S}{\theta_E},
\end{equation}

the root of the lens equation provides us the position of the distorted images,
\begin{equation}
u_\pm=(u\pm\sqrt{u^2+4})/2,
\end{equation}

Due to the conservation of surface brightness [53], the amplification of the background source is simply the ratio between the area of the images to the area of the source. So the amplification of the distorted images and the total amplification can be calculated by
\begin{equation}
A\pm=\big|\frac{u_\pm}{u}\frac{du_\pm}{du}\big|, \quad A=A_++A_-=\frac{u^2+2}{u\sqrt{u^2+4}}\sim\frac{1}{u},
\end{equation}

and yet it is only a function of u. This is the beauty of microlensing because one can calculate the light curve merely by the relative lens and source position projected onto the sky. If we assume the relative lens-source motion to be rectilinear, u can be decomposed into components parallel and perpendicular to the direction of the relative lens-source motion. u and A can thus be calculated as
\begin{equation}
A(t)=\frac{u(t)^2+2}{u(t)\sqrt{u(t)^2+4}},\quad u(t)=\sqrt{\big(\frac{t-t_0}{t_E}\big)^2+u_0^2},
  \end{equation}

where $t_0$ and $u_0$ are the time and impact parameter at the closest-approach. $t_E$ is the Einstein timescale, which is defined as the time required for the lens to traverse the Einstein radius
\begin{equation}
t_E=\frac{\theta_E}{\mu_{rel}}=0.214yr \big(\frac{M}{M_\odot}\big)^{1/2}\big(\frac{D_{OL}}{10kpc}\big)^1/2\big(1-\frac{D_{OL}}{D_{OS}}\big)^{1/2}\big(\frac{200 km/s}{V_{rel}}\big)
  \end{equation}

Since the first discovery of microlensing events in 1993 [9-11], thousands of events have been reported. However, the only parameter one can retrieve from the light curve is the event timescale tE. The Einstein timescale is unfortunately a degenerated parameter consisted of the lens mass, lens distance, and the relative lens-source velocity $\mu_{rel}$. Thus it is impossible to characterize the lens and the source of a single event through light curve measurement alone; the properties of the lens can only be revealed by statistic studies, unless special circumstances are present, e.g. parallax, finite source, binary lens caustic crossing and so forth.

\section{Astrometry comes to rescue}
As Gould (2000) [55] pointed out that, in order to break the microlensing degeneracy, one requires the measurements of both the angular Einstein radius $\theta_E$ and the microlens parallax
\begin{equation}
  \pi_E:=\frac{AU}{r_E}
  \end{equation}

where $r_E$ is the Einstein radius projected on the observer plane. The mass of the lens can be determined
without ambiguity [55]:

\begin{equation}
M=\frac{\theta_E}{\kappa \pi_E}, \quad \kappa=\frac{4GM}{c^2AU}\sim8.14\frac{mas}{M_\odot}.
\end{equation}

The microlens parallax can be derived from, for example, the Earth-orbital parallax caused by the orbital motion of Earth around the Sun [56]. This will result in parabolic lens-source trajectory instead of the rectilinear motion during the time of closest-approach in the geocentric observation. The information of microlens parallax can be obtained by fitting the tiny asymmetry in the light curve. In addition, the parallax information can also be derived if there are simultaneous observations at different locations, where the lens-source trajectories will appear to be different on the sky according to the observers’ locations, resulted in different shape of the microlensing light curves. Such effects are more prominent if the observers are farther away, for example, simultaneous observations from ground-based telescopes and from space telescopes [57].  Nevertheless for high magnification and fast moving lenses, it is also possible to detect the parallax effects even with different ground-based observatories [58]. The Einstein radius can also be obtained by 1) the finite-source effect; 2) the high resolution imaging; and 3) the astrometric trajectories. Besides the finite-source effect, the later two cases both rely on exquisite astrometry measurements of the microlensing events. Detail descriptions of each method are provided as follows. 

\subsection{Inferring the Einstein radius from finite-source effect}
When the lens transits the surface of the source during the course of microlensing, the point-source approximation is no longer valid. In this regard, we will have to integrate the magnification over the surface of the source by
\begin{equation}
A_{FS}(u|\rho_S)=\frac{\int_0^{2\pi} \int_0^{\rho_S}A\big[(u+r\mathrm{cos}\theta)^2+(r\mathrm{sin}\theta)^2\big]rdrd\theta}{\int^{2\pi}_0 \int^{\rho_s}_0 rdrd\theta}
  \end{equation}

where $\rho_S$ = $\theta_*/\theta_E$ is the angular source radius in units of $\theta_E$. When the lens is perfectly aligned with the source ($u$ = 0), the magnification will reduce to $(\rho_S^2+4)^{1/2}/\rho_S$, contrast to the infinite amplification in the point-source regime. We can thus derive the source size in units of the Einstein radius by fitting the light curve with one additional parameter ($\rho_S$). Finite-source effect in the microlensing thus serves as a powerful method to probe the surface-brightness profile of distant stars. Once we derive the source size in terms of the Einstein radius, we can further infer the Einstein radius by comparing $\rho_S$ to the actual source size derived from the empirical surface brightness – color relation. For example, Kervella et al. [59] proposed the following relation for angular diameter for A0-M2 dwarf stars or A0-K0 subgiants:

\begin{equation}
log(2\theta_*)=0.0755(V-K)+0.5170-0.2K
  \end{equation}

where V is in Johnson system and K with $\lambda$ = 2.0 – 2.4 $\mu$m. However, the typical value of $\theta_E$ is in the order of 0.5 mas while the $\theta_{*}$ is $\sim$ 0.5 $\mu$as, so we will need events with magnification larger than 1,000. The chance for the lens to transit the source is very slim, especially for the single lens cases, and we only have a handful of such single lens events observed so far. For binary/planetary lens events, however, if we detect multiple peaks in the microlensing light curves, this means that the lens are crossing the caustics, which leads to very high magnification. In this regard, we can often detect the finite-source effects from binary/planetary microlensing light curves, provided dense sampling around the light curve peaks.

\subsection{Measuring proper motion via high resolution imaging with HST or ground-based AO}

The first one is applicable when both the lens and the source are stellar objects, that is, to take a snapshot with very high precision astrometry long after the event. From the time span $\Delta$t and the separation between the lens and the source $\Delta\theta$ , one can easily calculate the relative lens-source velocity $\mu_{rel}$ . Combined with the Einstein timescale tE obtained from the light curve, one can thus derive the Einstein radius by $\theta_E=t_E\times\mu_{rel}$. So far there are only two such cases for single lens events, i.e. MACHO-LMC-5 [60] and MACHO-95-BLG-37 [61], because this method requires the lens-source relative velocity to be very large and both the lens and the source must be luminous enough for detection. For the case of MACHO-LMC-5, the parallax effects can be inferred with tiny asymmetry in the light curve, which sheds light on the properties of the lens, such as its mass and location (Alcock et al., 2001; Gould, 2004; Drake et al., 2004; Gould et al., 2004) [56, 60,62-63].

For the case of binary/planetary microlensing, resolving both the lens and the source not only provides constraints on the Einstein radius, but also helps pin down the nature of the secondary / planetary companions [64]. This is because from the light curve modeling we can only obtain the mass ratio between the primary and secondary lens, and the flux from the primary lens provides us additional constraint to better inferring the mass of the planetary companion. In addition, as microlensing experiments are targeting very crowded stellar fields, high resolution imaging can also probe the flux contamination from objects that are unrelated to the lens and source. This is important to interpret the flux excess on top of the source from light curve modeling, because often it is assumed that the flux excess comes solely from the lens. However, recent high resolution observations suggest that unrelated objects in the vicinity of the microlensing events can also contribute to the flux access [65], especially with the coarse pixel resolution delivered by ground-based microlensing surveys. While there have been numerous HST and/or AO follow-up of planetary microlensing events, only OGLE-2005-BLG-169 shows resolved lens and source with HST [66] or ground-based AO [65]. Nevertheless it is important to obtain several epochs of high resolution observations, especially several years after the light curve maximum to resolve the lens and the source. Thus we provide a summary of high resolution imaging on previous events in Table 1, to encourage further follow-up in the future.

\begin{table}[H]
\caption{High resolution observations of microlensing events.}
\centering
\begin{tabular}{lll}
\toprule
\textbf{Planetary events} & \textbf{High-res. Obs.} & \textbf{Reference}\\
\midrule
OGLE-2003-BLG-235 & HST/ACS on 2006-5-1 & [67]\\
OGLE-2005-BLS-071 & HST/ACS on 2005-5-23, 2006-2-21 & [68]\\
OGLE-2005-BLG-169 & HST/WFC on 2011-10-19, 2012-2-16, & [65-66] \\
& 2012-2-22, 2012-2-23, & \\
& Keck/NIRC2+AO on 2013-7-18 & \\
MOA-2007-BLG-192 & VLT/NACO+AO on 2007-9-6, 2009-7-2, & [69] \\
& 2009-7-23 & \\
OGLE-2007-BLG-349 & HST/WFPC2 and NICMOS on 2007-10-8, & [70] \\
& 2008-5-15, & \\
& VLT/NACO+AO on 2007-10-13, 2008-8-4, & \\
& 2008-8-8 & \\
OGLE-2007-BLG-368 & Keck/NIRC2+AO on 2007-8-20 & [71]\\
MOA-2008-BLG-310 & VLT/NACO+AO on 2008-7-28 & [72] \\
MOA-2011-BLG-293 & Keck/NIRC2+AO on 2012-5-13 & [64]\\
OGLE-2012-BLG-563 & Subaru/IRCS+AO on 2012-7-28 & [73]\\
OGLE-2012-BLG-0950 & Keck/NIRC2+AO on 2012-7-18 & [74] \\
OGLE-2012-BLG-0026 & Keck/NIRC2+AO on 2012-5-6 & [75] \\
& Subaru/IRCS+AO on 2012-7-28 & \\
MOA-2016-BLG-227 & Keck/NIRC2+AO on 2016-8-13 & [76] \\
\midrule
\textbf{Non-planetary events} & \textbf{High-res. Obs.} & \textbf{Reference}\\
\midrule
MACHO-LMC-1 & HST/WFPC2 on 1997-12-16 & [77] \\
MACHO-LMC-4 & HST/WFPC2 on 1997-12-12 & [77] \\
MACHO-LMC-5 & HST/WFPC2 on 1999-5-13 & [77] \\
MACHO-LMC-6 & HST/WFPC2 on 1999-8-26 & [77] \\
MACHO-LMC-7 & HST/WFPC2 on 1999-4-12 & [77] \\
MACHO-LMC-8 & HST/WFPC2 on 1999-3-12 & [77] \\
MACHO-LMC-9 & HST/WFPC2 on 1999-4-13 & [77] \\
MACHO-LMC-13 & HST/WFPC2 on 2000-7-28 & [78] \\
MACHO-LMC-14 & HST/WFPC2 on 1997-5-13 & [77] \\
MACHO-LMC-15 & HST/WFPC2 on 2000-7-17 & [78] \\
MACHO-LMC-18 & HST/WFPC2 on 2000-7-21 & [78] \\
MACHO-LMC-20 & HST/WFPC2 on 2000-7-29 & [78] \\
MACHO-LMC-21 & HST/WFPC2 on 2000-7-26 & [78] \\
MACHO-LMC-23 & HST/WFPC2 on 2000-7-18 & [78] \\
MACHO-LMC-25 & HST/WFPC2 on 2000-7-14 & [78] \\
\bottomrule
\end{tabular}
\end{table}

\subsection{Detecting astrometric trajectory during the course of microlensing}
Another way to determine the Einstein radius is through astrometric microlensing. The thoretical grounds were laid down in 1990's [79-82]. Paczynski (1996, [83]) was the first one to bring the idea forth by estimating the probability of observing such phenomena. Boden et al. (1998, [84]) further considered astrometric observations with expected errors of planned space experiments. The most extensive works on astrometric microlensing were provided by Dominik \& Sahu (2000, [85]), who not only provided a thorough review of astrometirc microlensing of stars, but also predicted the event rate by $SIM$ and $Gaia$.
  The idea of astrometric microlensing is that, although the state-of-art observatories are not able to resolve the two microlensed images, it is possible to measure the astrometric centroid of the plus- and minus-image relative to the source,
        \begin{equation}
\delta\theta_C=\frac{A_+\theta_++A_-\theta_-}{A_++A_-}-u = \frac{u}{u^2+2}
        \end{equation}

with maximum deviation $\sim$ 0.35$\theta_E$ occurs at $u$ = $\sqrt{2}$. It has been shown that the astrometric centroid relative to the source will trace out an ellipse, and the size of of the ellipse gives the scale of the Einstein radius, as shown in Fig. 2. This implies that we can determine the Einstein radius for every single event if the astrometric signal is large enough to be observed. For example, a source in the Galactic bulge lensed by an object of 0.5 M$_\odot$ on the half way has $\theta_E$ $\sim$ 0.7 mas. This astrometric accuracy is achivable with $Gaia$, albeit limited only to brighter stars (Gaia Collaboration 2016); for astrometric microlensing the per-observations astrometric measurements will be needed and these will be able to detect Einstein radii only for most massive lenses producing $\theta_E$ of few mas. However, the astrometric signal for self-lensing events towards Magellanic Clouds and M31 will be beyond the detection limit of $Gaia$.

\begin{figure}[H]
\centering
\includegraphics[width=15 cm]{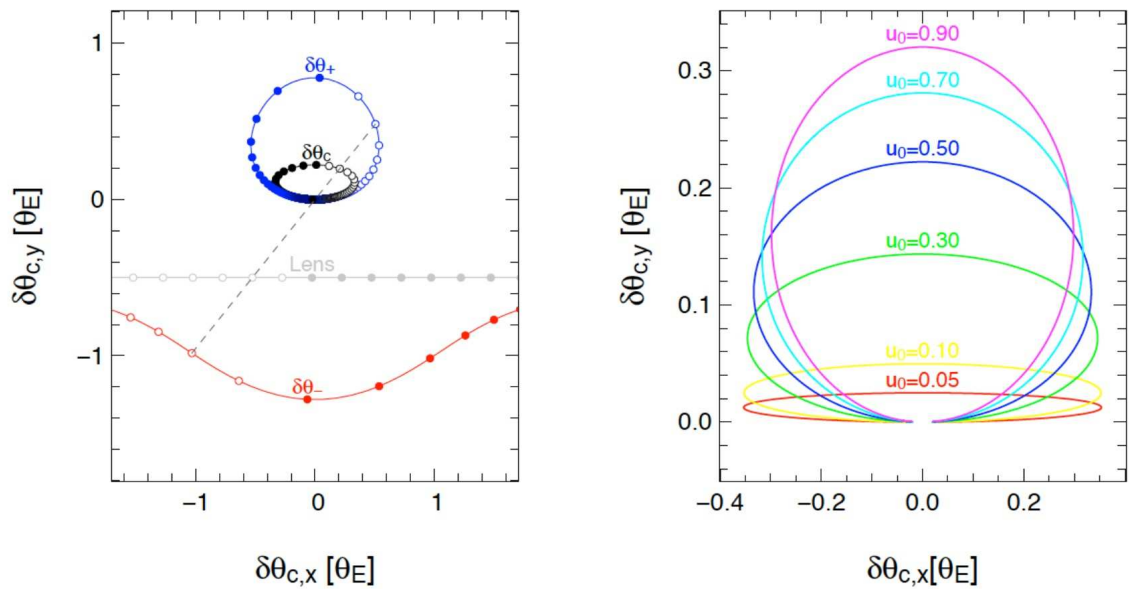}
\caption{Centroid shifts of a single lens microlensing event. Left: trajectories of the plus-image (blue), minus-image (red), and the centroid (black). The lens trajectory is shown in grey. We assume $t_0$ = 0, $t_E$ = 10 days and $u_0$ = 0.5 $\theta_E$. Right: centroid shifts for different values of $u_0$. This figure is adopted from ``Lee et al., MNRAS 2010, 407, 1597-1608''.}
\end{figure}

As the Einstein radius is in proportion to the lens mass, it is easier to detect astrometric microlensing signals from massive lenses, especially black holes. In this regard, there have been several attempts to observe astrometric microlensing events with either ground-based AO [86] or satellite [87]. However, there have not been confirmative detections of astrometric microlensing signals for black holes for the moment, though it is possible to turn the null detection into constrains on the mass of the black holes. As the time being, the only affirmative detection of astrometric microlensing is from a nearby white dwarf Stein 2051B [88]. Instead of following-up on-going microlensing events, Sahu et al. made use of nearby, high proper motion stars and white dwarfs, and selected the ones that will pass very close to a background star for high precision astrometry follow-up. During the selection process, they found out that Stein 2051B, the 6th closest white dwarf, would pass a 19.5 magnitude background star in March 2014, with an impact parameter of 0.1 arcsec. They used HST/WFC3 to obtain 8 epochs of observations between October 2013 and November 2015. Though the background source is 400 times fainter than Stein 2051B, they were nevertheless able to extract astrometric microlensing signal at a few mas level with $>$ 10 sigma detection delivered by HST/WFC3. With the astrometric microlensing, they were able to determine the Einstein radius caused by Stein 2015B, and further pin down its mass, which is in good agreement with the mass-radius relation of white dwarf.

\section{Prospects}
While resolving the source and lens, as well as measure the astrometric microlensing are feasible with the state-of-the-art instruments, in order to better break the microlensing degeneracy, it is essential to achieve exquisite astrometry. For example, Proft et al. [89] predicted 43 astrometric microlensing events will be caused by high proper motion stars between 2012 and 2019. However, the majority of them will have centroid shifts below 0.1 mas, and only nine of them will have measurable centroid shift between 0.1 and 4 mas. Among these nine candidates, the most promising event (largest centroid shift) is Stein 2051B, which indeed have been measured by HST. If we can increase the astrometry accuracy of HST, e.g. with the spatial scanning technique, we will be able to detect the astrometric microlensing signals from other events.

 However, most of the microlensing events are discovered by high-cadence photometry surveys of the Galactic bulge or Magellanic Clouds, where the lens are faint and it is not possible to predict the timing of the microlensing in priori. In this regard, all-sky astrometry satellites e.g. $Gaia$ will be essential to provide a comprehensive astrometric measurements for the vast majority of the microlensing events. $Gaia$ is now surveying the whole sky; by the end of its 5 year mission, $Gaia$ will deliver multi-epoch (~80 epochs), sub-milli arcseconds astrometric measurements. This will enable the mass determination of numerous microlensing events, both predicted by Proft et al. [89] with known high proper motion stars as lenses, and events with unseen lenses that will be continuously discovered by the on-going microlensing surveys, such as OGLE, MOA, and KMT. After $Gaia$, we will still be able to obtain exquisite astrometry with forth-coming observatories, especially ground-based AO with 30-meter class telescopes. Such measurements will provide essential measurements for the astrometric microlensing, providing direct measurements or stringent constraints on the mass of isolated, stellar mass black holes. 

\acknowledgments{We would like to thank the referees for their insightful comments which greatly improved the manuscript.}


\conflictsofinterest{The authors declare no conflict of interest.}





\end{document}